\documentclass[final,5p]{elsarticle}
\usepackage{lineno,hyperref}
\usepackage{amsmath}

\journal{Physica B: Condensed Matter}

\begin{document}

\begin{frontmatter}
\title{Correlation effects in superconducting quantum dot systems}

\author[addFZU,addMFF]{Vladislav Pokorn\'y\corref{ca}}
\cortext[ca]{Corresponding author}
\ead{pokornyv@fzu.cz}

\author[addMFF]{Martin \v{Z}onda}
\ead{martin.zonda@karlov.mff.cuni.cz}

\address[addFZU]{Institute of Physics, The Czech Academy of Sciences, Na Slovance 2, CZ-18221 Praha 8, Czech Republic}
\address[addMFF]{Department of Condensed Matter Physics, Faculty of Mathematics and Physics, 
Charles University in Prague, Ke Karlovu 5, CZ-12116 Praha 2, Czech Republic}

\begin{abstract}
We study the effect of electron correlations on a system consisting of a 
single-level quantum dot with local
Coulomb interaction attached to two superconducting leads. We use the
single-impurity Anderson model with BCS superconducting baths to study the interplay 
between the proximity induced electron pairing and the local Coulomb interaction. 
We show how to solve the model using the continuous-time hybridization-expansion quantum Monte
Carlo method. The results obtained for experimentally relevant parameters 
are compared with results of self-consistent second order
perturbation theory as well as with the numerical renormalization group method. 
\end{abstract}


\end{frontmatter}

\section{Introduction}
Conventional Josephson junctions had become a standard building blocks of various electronics devices
including SQUIDs \cite{W1996}, RSFQs \cite{LS1991}, and qubits \cite{CW2008} 
in quantum computing. 
No wonder that their tunable generalizations, the superconducting quantum dots, gain a lot of attention 
from both theorist and experimentalist. 
These hybrids, where a quantum dot is placed between two superconducting leads, promise a great deal of 
technological advances such as quantum supercurrent  transistors \cite{Herrero-2006},
monochromatic single-electron sources \cite{Zanten-2016} or single-molecule SQUIDs \cite{Cleuziou-2006}. 
Of no less importance is the fact that they are rich and relatively easy to deal with playgrounds for 
studying various physical phenomena. These include the competition between Kondo effect and 
superconductivity \cite{Domanski-2017}, the Andreev subgap transport \cite{Rodero-2011} 
as well as quantum phase transitions from impurity spin-singlet to spin-doublet ground 
state which are in the experiments signaled by the sign reversal 
of the supercurrent (0-$\pi$ transition) and accompanied by a crossing of the subgap 
Andreev bound states (ABS) \cite{PJZG2013,ZPJN2015,Delagrange-2016}.

The superconducting quantum dot can be adequately described by a single impurity 
Anderson model (SIAM) coupled to BCS leads \cite{Luitz-2012}. 
Consequently, a lot of different theoretical approaches have been applied 
to study this system. Many important results have been obtained using
various (semi)analytical methods based on different perturbation
approaches \cite{Rodero-2011,Clerk-2000,Karrasch-2008,Meng-2009}.
Moreover, as was shown in recent studies \cite{ZPJN2015,ZPJN2016}, a surprisingly 
large portion of the parametric space of the superconducting SIAM 
can be reliably covered with a properly formulated second-order perturbation theory (2ndPT) 
in the on-dot electron Coulomb interaction. Unfortunately,
this method cannot describe the $\pi$-junction behavior due to the
ground-state degeneracy.

None of the mentioned (semi)analytical perturbative methods can cover all 
experimentally relevant cases. Therefore there is a big demand for ``heavy'' numerical methods.
A very good quantitative agreement with the experiments can be obtained with the numerical 
renormalization group (NRG) \cite{Yoshioka-2000,PJZG2013} and quantum Monte Carlo 
(QMC) \cite{la2010,Luitz-2012} methods.
Although both methods have some disadvantages, including their computational demands,
they have, besides parametric universality, one big practical advantage.
Namely, the existence of well-tested versatile open-source software packages.

In the present paper we focus on the continuous-time hybridization-expansion quantum Monte
Carlo (CT-HYB) \cite{qmc-RMP2011} implementation for experimentally inspired parameters \cite{PJZG2013} 
representing a strong coupling regime, which is beyond the reach for most 
(semi)analytical techniques.

We show how to include the superconductivity into CT-HYB quantum Monte Carlo solver. 
Then we study various single-particle quantities as functions of the gate voltage. 
We discus how they behave near quantum phase transition and
show that the CT-HYB can be reliably used to obtain the phase diagram. 
We also use numerical analytical continuation to obtain the spectral function. 
We compare all obtained CT-HYB results with either 2ndPT or the NRG.      

\section{The model Hamiltonian}
We describe the system by the single-impurity Anderson model with BCS leads. 
The Hamiltonian reads
\begin{equation}
\label{Ham}
\mathcal{H}=\mathcal{H}_{dot}+\sum_{s=L,R}(\mathcal{H}^s_{lead}+\mathcal{H}^s_c)
\end{equation}
where $s=L,R$ denotes the left and right leads. The impurity Hamiltonian describes a single-level atom with
local Coulomb repulsion $U$ and on-site energy $\varepsilon_\sigma=\varepsilon+\sigma B$, 
where $B$ is the external magnetic field
\begin{equation}
\mathcal{H}_{dot}=\sum_\sigma\varepsilon^{\phantom{\dag}}_\sigma d_\sigma^\dag d_\sigma^{\phantom{\dag}}
+Ud_\uparrow^\dag d_\uparrow^{\phantom{\dag}} d_\downarrow^\dag d_\downarrow^{\phantom{\dag}}.
\end{equation}
The Hamiltonian of the BCS leads reads
\begin{equation}
\begin{split}
 \mathcal{H}^s_{lead}&=\sum_{\mathbf{k}\sigma}
\varepsilon(\mathbf{k})c_{s,\mathbf{k}\sigma}^\dag c_{s,\mathbf{k}\sigma}^{\phantom{\dag}} \\
&-\Delta\sum_\mathbf{k}(e^{i\Phi_s}
c_{s,\mathbf{k}\uparrow}^\dag c_{s,\mathbf{-k}\downarrow}^\dag+\textrm{H.c.})
\end{split}
\end{equation}
where $\Delta e^{i\Phi_s}$ is the complex gap parameter. We assume the same gap size in both leads, 
$\Delta_L=\Delta_R=\Delta$, meaning that the leads are made from the same material, as it is usual in the experimental setups.
Finally, the coupling part reads
\begin{equation}
\mathcal{H}^s_c=-\sum_{\mathbf{k}\sigma}
t_s(c_{s,\mathbf{k}\sigma}^\dag d_\sigma^{\phantom{\dag}}+\textrm{H.c.})
\end{equation}
where $t_s$ denotes the tunneling matrix element.

Hamiltonian~\eqref{Ham} does not conserve the electron number and therefore cannot be solved directly 
using standard CT-HYB technique.
To circumvent this problem we utilized a canonical particle-hole transformation in the spin-down sector
\begin{equation}
\label{eq:trans}
\begin{split}
d^\dag_\uparrow&\rightarrow d^\dag_\uparrow,\quad 
d^\dag_\downarrow \rightarrow d^{\phantom{\dag}}_\downarrow,\quad
d^{\phantom{\dag}}_\uparrow \rightarrow d^{\phantom{\dag}}_\uparrow,\quad 
d^{\phantom{\dag}}_\downarrow \rightarrow d^\dag_\downarrow,\\
c_{\mathbf{k}\uparrow}^\dag&\rightarrow c_{\mathbf{k}\uparrow}^\dag,\quad
c_{\mathbf{k}\downarrow}^\dag\rightarrow c^{\phantom{\dag}}_{\mathbf{-k}\downarrow},\quad
c^{\phantom{\dag}}_{\mathbf{k}\uparrow}\rightarrow c^{\phantom{\dag}}_{\mathbf{k}\uparrow},\quad
c^{\phantom{\dag}}_{\mathbf{k}\downarrow}\rightarrow c_{\mathbf{-k}\downarrow}^\dag,
\end{split}
\end{equation}
previously used by Luitz and Assaad \cite{la2010} to include superconductivity in the continuous-time 
interaction-expansion (CT-INT) QMC calculations. 
The new quasiparticles are identical to electrons in the spin-up sector and to holes in the spin-down
sector. This transformation maps our system to SIAM with attractive interaction $-U$ and off-diagonal 
hybridization of the quantum dot with the leads. 
The local energy levels transform as $\varepsilon_\sigma \rightarrow \sigma\varepsilon_\sigma$. 
Since $\varepsilon_\sigma=\varepsilon+\sigma B$ and 
$\sigma^2=1$, this transformation maps the local energy $\varepsilon$ on the magnetic field $B$ and vice versa.
The dispersion and tunneling matrix elements transform in the same manner, 
$\varepsilon(\mathbf{k}) \rightarrow \sigma\varepsilon(\mathbf{k})$ and $t_s\rightarrow\sigma t_s$.
The resulting Hamiltonian conserves the total electron number and can be solved using standard CT-HYB 
implementations.

\section{The CT-HYB method}
We use the TRIQS/CTHYB Monte Carlo solver \cite{cthyb2016,triqs2015}. 
We consider a flat density of states in the leads of finite half-width $D=30\Delta$. The coupling of the quantum dot to the leads
is described by tunneling rates $\Gamma_s=\pi|t_s|^2/(2D)$.
We denote $\Gamma=\Gamma_R+\Gamma_L$ and consider only the symmetric coupling $\Gamma_R=\Gamma_L$. 
Any asymmetric coupling $\Gamma_R\neq\Gamma_L$
with the same total $\Gamma$ can be easily gained from the symmetric solution using a simple analytical 
relation derived in Ref. \cite{KZN2017}.

Continuous-time quantum Monte Carlo belongs to a family of inherently finite-temperature methods
and the calculations are usually restricted to rather high temperatures.
However, since the typical energy scale in our setup is the superconducting gap 
$\Delta\sim 100\mu$eV, it allows us to easily reach experimental range of temperatures $T\sim10-100$mK.

The biggest disadvantage of CT-HYB in comparison with NRG or 2ndPT is 
that the calculation is performed on the imaginary-time axis. 
Obtaining the spectral function from imaginary-time data is a well-known ill-defined problem. 
Various numerical methods are used to perform the analytic continuation, the maximum entropy method being the most 
common one \cite{JG1996}. However, this method fails to resolve sharp spectral features like the Andreev bound states.
Therefore we use the Mishchenko's stochastic optimization method (SOM) \cite{MPSS2000} in its 
recent implementation \cite{SOM-code} which is better suited to our needs.

\section{Results}
Our calculations are inspired by the experiment of Pillet et.~al. \cite{PJZG2013}. 
The paper describes the tunneling spectroscopy measurement
performed on a carbon nanotube connected to superconducting aluminum leads. 
Experimental results show the Andreev bound states as functions of gate voltage 
and are nicely reproduced using the NRG method. 
The superconducting gap is $\Delta=150\mu$eV, Coulomb interaction $U\approx2$meV 
and the phase difference is zero ($\Phi_L=\Phi_R$). We use these parameters in our calculations
and set the magnetic field $B$ to zero. It is worth to note that we did not encounter any 
fermionic sign problem during the calculation.

\begin{figure}[h]
\hspace{5mm}
\includegraphics[width=7cm]{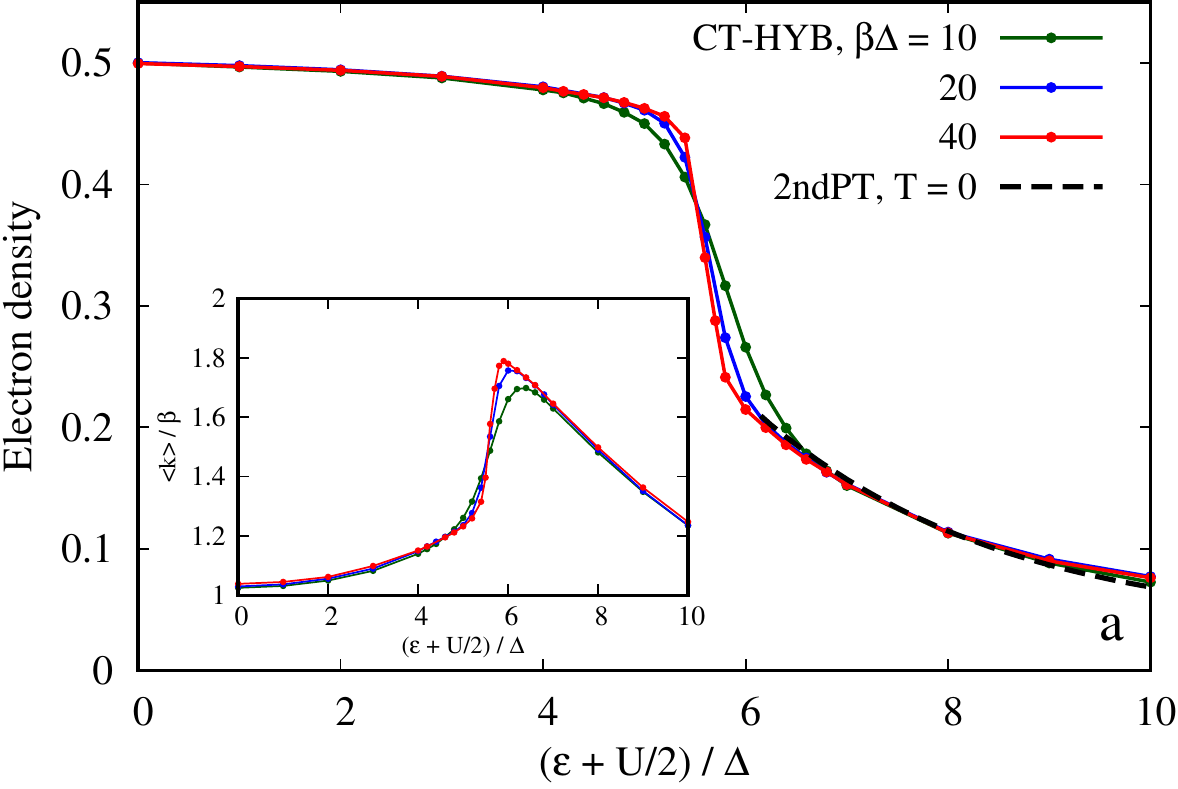}\\

\hspace{5mm}
\includegraphics[width=7cm]{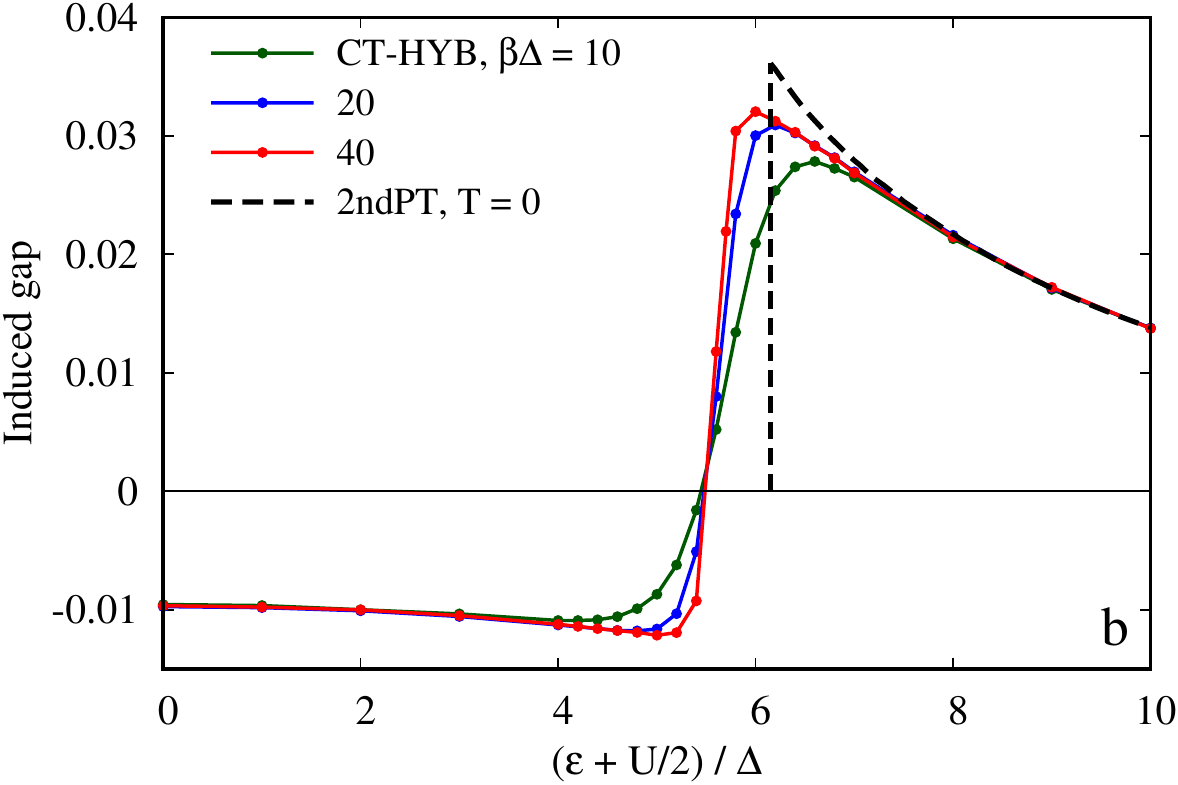}
\caption{The diagonal (electron density, panel \textbf{a}) 
and off-diagonal (induced gap, panel \textbf{b}) part of the occupation 
matrix as functions of the local energy level $\varepsilon$. 
Solid lines: CT-HYB quantum Monte Carlo results for three values of 
inverse temperature $\beta\Delta=10$ (green), $20$ (blue) and $40$ (red) that correspond to temperatures 
$T=175$~mK, $77$~mK and $44$~mK, respectively. Dashed black line: 2ndPT result
at zero temperature (available only in the $0$-phase). 
Inset: Average perturbation order $\langle k\rangle$ of the CT-HYB 
calculation scaled by the inverse temperature $\beta$. 
Line colors in the inset correspond the main plots.}
\label{Fig1}
\end{figure}

In Fig.~\ref{Fig1} we plot the diagonal (panel \textbf{a}) and the off-diagonal (panel \textbf{b}) 
part of the occupation matrix as functions of the shifted local energy level 
$\varepsilon_U=\varepsilon+U/2$ ($\varepsilon_U=0$ represents the half-filled dot). 
From now on we use $\Delta$ as the energy unit.
We restrict ourselves to positive values of $\varepsilon_U$ as the rest can be determined from symmetry.
We chose parameters $U=13.3\Delta$ and $\Gamma_L=\Gamma_R=0.45\Delta$ which are within the experimental range. 
The three solid lines are CT-HYB results calculated at inverse
temperatures $\beta\Delta=10$ (green), $20$ (blue) and $40$ (red) that correspond to 
temperatures $T=175$~mK, $77$~mK and $44$~mK, respectively.
The diagonal part corresponds to the electron density $n=\langle d^{\dag}d\rangle$. It varies very weakly in 
$\pi$ phase, then changes abruptly at the phase transition. 

The position of the phase transition can be more easily
determined from the off-diagonal part, which represents the induced gap $\mu=\langle d^{\dag}d^{\dag}\rangle$. 
This parameter is negative in the $\pi$-phase and positive in the $0$-phase.
We see that the temperature has very little effect on the position of the phase transition 
which takes place at $\varepsilon_U\approx 5.5 \Delta$.
We included also the results of the 2ndPT method, which is available only in the $0$-phase. 
It fits very well the CT-HYB results in this phase although it gives the phase 
transition at $\varepsilon_U=6.15\Delta$ (c.a. $12\%$ error). However, this discrepancy is expected
as we are investigating a strong coupling regime ($U/\Delta \gg 1$ and $U/\Gamma \gg 1$).

The inset of panel \textbf{a} shows the average perturbation order $\langle k\rangle$ 
scaled by the inverse temperature $\beta$ for the CT-HYB calculations in the main panels. 
This quantity is an estimator of the kinetic energy \cite{H2007}.
It scales linearly with $\beta$ and exhibits a maximum just above the phase transition point. 
Although the zero temperature extrapolation of the position of maxima could be in principle used 
to estimate the phase transition point, safe determination of its position from $\langle k\rangle$ 
requires a rather elaborate procedure \cite{HWWW2016}.

\begin{figure}[h]
\hspace{5mm}
\includegraphics[width=7cm]{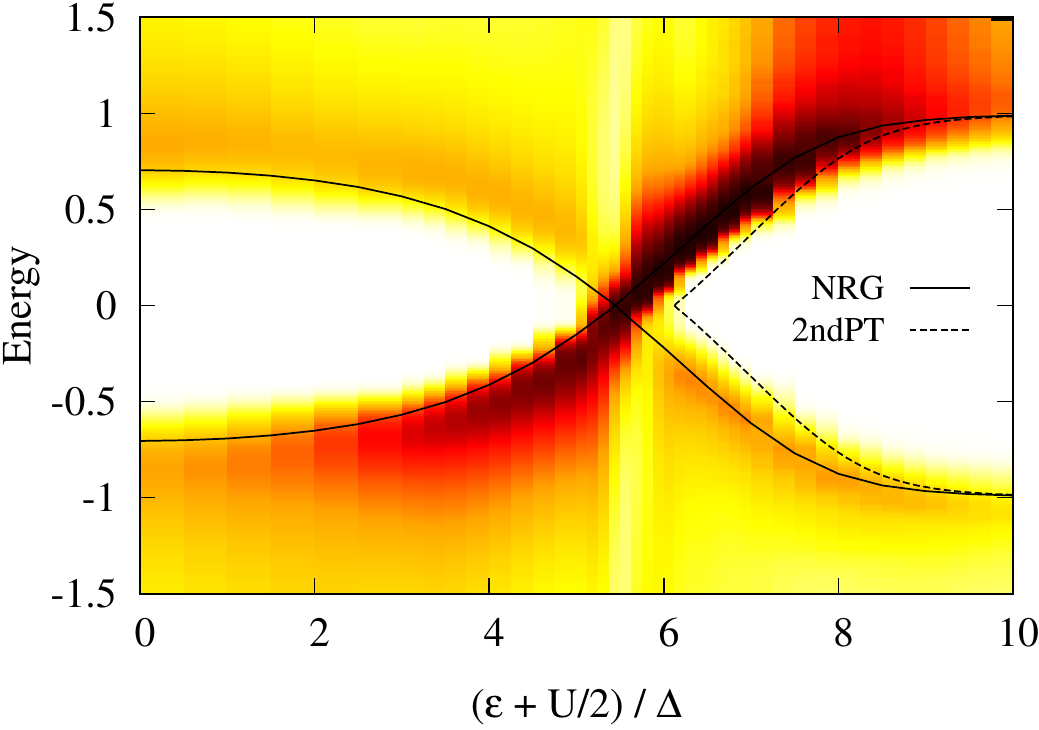}
\caption{Color map of the spectral function around the gap region for the same parameters 
as in Fig.~\ref{Fig1} calculated using the stochastic optimization method from CT-HYB 
results at inverse temperature $\beta\Delta=40$ (corresponds to $T=44$~mK).
Black solid and dashed lines are the zero-temperature NRG and 2ndPT results, respectively. 
Energy on the vertical axis is in units of $\Delta$.}
\label{Fig2}
\end{figure}

Calculation of a spectral function requires much more precise QMC data than the calculation of 
an expectation value due to the underlying, ill-defined analytic continuation procedure. While the 
expectation values with reasonable error bars can be obtained within few CPU-hours, 
calculation of a spectral function, including the SOM procedure, 
takes usually more than 100, depending strongly on the temperature and the coupling strength $\Gamma$.
In Fig.~\ref{Fig2} we plot the color map of the spectral function at $\beta\Delta=40$ ($T=44$mK) 
in the vicinity of the gap region as it can be directly compared to the experimental data. 
We use the same parameters as in Fig.~\ref{Fig1} and compare it with the 
position of ABS calculated using NRG and 2ndPT methods at $T=0$. 
NRG results were obtained using NRG Ljubljana code \cite{Ljubljana-code}. 
The ingap maxima of the spectral function in the $0$-phase and around the transition point 
are in very good agreement with the positions of ABS calculated by NRG. 
In the $\pi$-phase the position of the maxima tends to shift to higher energies. 
This is surprising as the electron density and the induced gap values are in good 
agreement with NRG even in this region. The position of the peaks also
does not depend on the temperature and while it does
depend on the width of the non-interacting band $D$, this dependence is weak and it effects 
the position of the maxima equally in $0$ and $\pi$ phase, therefore it cannot explain this
discrepancy.

\begin{figure}[h]
\hspace{5mm}
\includegraphics[width=7cm]{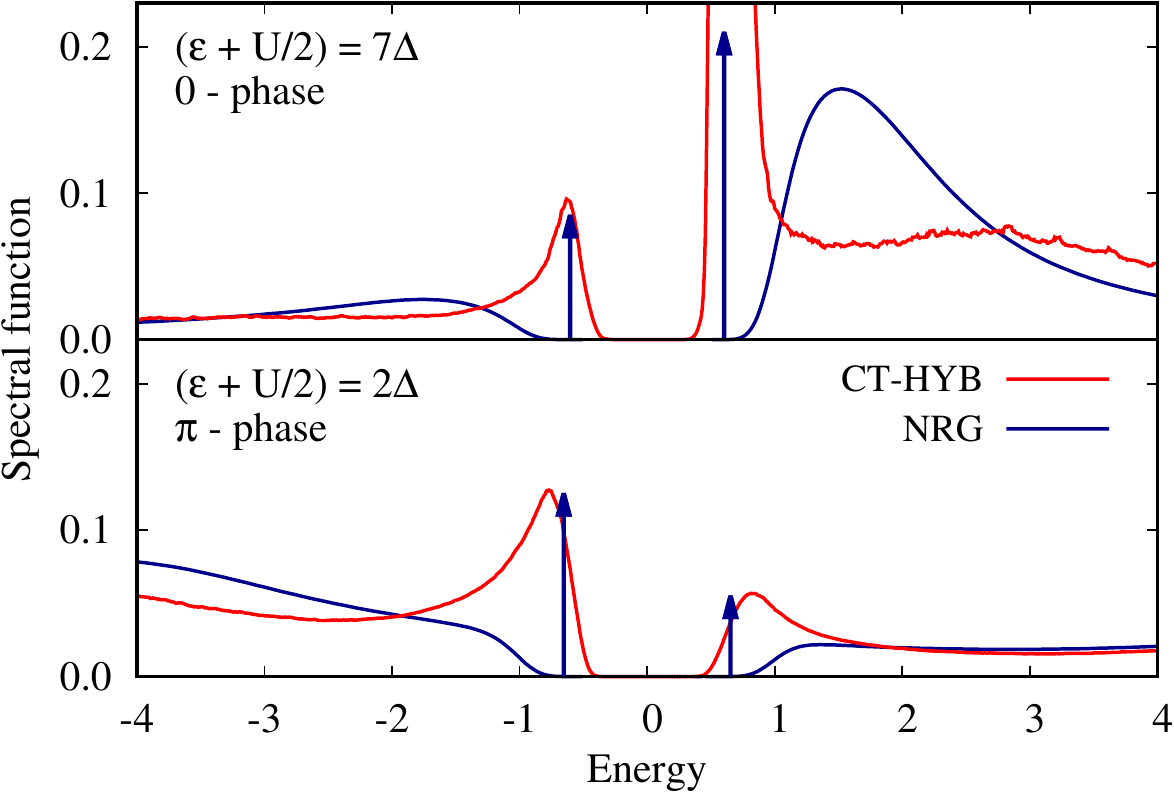}
\caption{Comparison of the spectral functions obtained using NRG (blue lines) and CT-HYB (red lines) for two
values of the local energy level $\varepsilon$. Arrows represent positions of ABS from NRG calculation.
CT-HYB calculation was performed at inverse temperature $\beta\Delta=40$. The NRG curves were calculated at zero temperature
where we used a log-Gaussian broadening of the continuum states with broadening parameter set to 0.2, see Ref.~\cite{Ljubljana-code}.
Energy on the horizontal axis is in units of $\Delta$.}
\label{Fig3}
\end{figure}

In order to get some insight into this problem we plotted in Fig.~\ref{Fig3}
the spectral functions calculated using NRG and CT-HYB methods. The top panel shows results for $\varepsilon_U=7\Delta$
which is in $0$-phase. We see that the arrows that represent ABS from NRG calculation match the maxima of the 
spectral function obtained using SOM procedure from CT-HYB data. We also see that CT-HYB spectra are missing the
structure just above the gap edges at $\pm\Delta$. 
Bottom panel shows spectral functions for $\varepsilon_U=2\Delta$ ($\pi$-phase). 
The mismatch between the arrows and the maxima is clearly visible and
we do not have a satisfactory explanation of this discrepancy.

\begin{figure}[h]
\hspace{5mm}
\includegraphics[width=7cm]{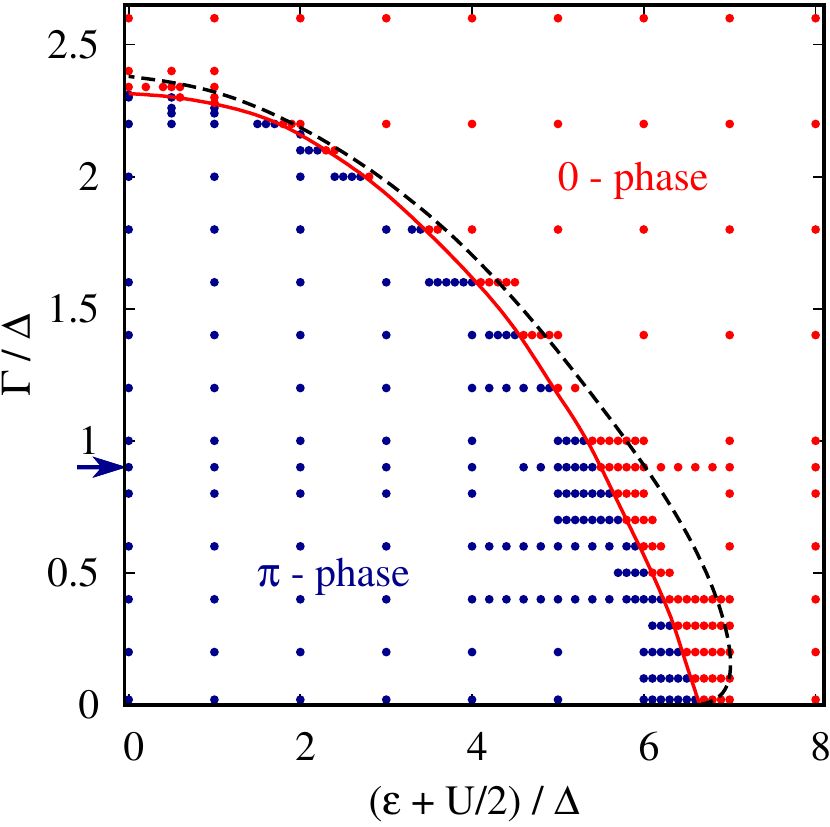} 
\caption{Phase diagram in the $\varepsilon-\Gamma$ plane. Red solid line: phase
boundary calculated using CT-HYB at inverse temperature $\beta\Delta=20$ 
(corresponds to $T=77$~mK). Black dashed line:
phase boundary calculated using 2ndPT at zero temperature. The
blue arrow marks the cut at $\Gamma=0.9\Delta$ along which the data in Figs.~\ref{Fig1} and~\ref{Fig2} are plotted. 
Red and blue points represent individual QMC measurements.}
\label{Fig4}
\end{figure}

The local energy level $\varepsilon$ is a parameter that can be easily tuned 
in the experimental setups by changing the gate voltage. 
On the other hand, the tunneling rate $\Gamma$ is very hard to measure and it is
usually obtained from numerical fits \cite{PJZG2013}. Studying the relation between these parameters
is therefore important for interpretation of the experimental results. 
In Fig.~\ref{Fig4} we plotted the phase diagram in 
the $\varepsilon_U-\Gamma$ plane. Red solid line represents the phase
boundary calculated using CT-HYB at inverse temperature $\beta\Delta=20$ that corresponds to $T=77$~mK. 
This boundary was determined from the positivity of the induced gap $\mu$. This line does not change 
with further decreasing temperature beyond the resolution of the plot.
We also included 2ndPT result for comparison. As this perturbation expansion is performed in 
$U/\Gamma$ parameter it differs more for small 
values of $\Gamma$ where it develops a ``hump'' as already pointed out in 
Ref. \cite{ZPJN2016}. The two lines then 
meet for $\Gamma=0$ at the exact result $\varepsilon_U=U/2$.
The blue arrow marks the cut at $\Gamma=0.9\Delta$ along which the data in 
Figs.~\ref{Fig1} and~\ref{Fig2} are plotted.

\section{Conclusions}
We studied a $0-\pi$ quantum phase transition in a single-level quantum dot connected to two superconducting BCS 
leads using the continuous-time hybridization-expansion quantum Monte-Carlo method. 
We used the $U$ and $\Delta$ parameters from experiment described in Ref. \cite{PJZG2013} 
in order to stay in a realistic region of the parameter space. 
Performing an electron-hole transformation in the spin-down sector we mapped the system on a model 
that can be solved using CT-HYB method as implemented in the TRIQS package. We presented results
as functions of the gate voltage $\varepsilon$ as this parameter is easily tunable in the experiment.
We showed how the $0-\pi$ quantum phase transition point can be extracted from the behavior of the induced gap
and presented the finite-temperature spectral function as well as the phase diagram in the $\varepsilon-\Gamma$ plane
that can be used to determine the value of the tunneling rate $\Gamma$.

In summary, we showed that CT-HYB is an effective method for studying superconducting 
quantum dot systems, where the interaction strength is the dominant energy scale. 
The present formulation is sign problem free and one can access the low-temperature
region using reasonable amount of computational resources. We also showed how the spectral function
can be obtained using analytic continuation based on the Mishchenko's stochastic sampling method
in order to study the behavior of the subgap Andreev bound states. 
Comparing the position of the subgap maxima with ABS frequencies from the NRG calculation shows
good agreement in the $0$-phase but a discrepancy in the $\pi$-phase which is of unknown origin.
Furthermore, the model can be generalized to include a normal (non superconducting) electrode.
As already pointed out in Ref. \cite{Domanski-2017} where such a three-terminal device was studied, 
NRG and 2ndPT methods fail in this setup except special cases and CT-HYB becomes the method of choice.

\section*{Acknowledgments}
Research on this problem was supported by Grant No. 15-14259S of the Czech Science Foundation (V.P.)
and the Grant No. DEC-2014/13/B/ST3/04451 of the National Science Centre (Poland) (M.\v{Z}.).
Access to computing and storage facilities owned by parties and projects contributing to the 
National Grid Infrastructure MetaCentrum provided under the programme ``Projects of Large 
Research, Development, and Innovations Infrastructures'' (CESNET LM2015042), is greatly appreciated.
V.~P. thanks Roberto Mozara for the help with the stochastic optimization method.


\end{document}